\documentclass[longauth]{aa} 

\usepackage{floatrow}
\usepackage{siunitx}
\usepackage{dcolumn}
\usepackage{xcolor}

\usepackage{graphicx}
\usepackage{txfonts}
\usepackage[normalem]{ulem}
\newcommand\Tstrut{\rule{0pt}{2.6ex}}       
\newcommand\Bstrut{\rule[-0.9ex]{0pt}{0pt}} 
\newcommand{\TBstrut}{\Tstrut\Bstrut} 

\begin{document} 
\title{Upper Limits on Very-High-Energy Gamma-ray Emission from Core-Collapse Supernovae Observed with H.E.S.S.}
\titlerunning{UL on core-collapse SNe observed with H.E.S.S.}	
\authorrunning{H.~Abdalla et al}
\makeatletter
\renewcommand*{\@fnsymbol}[1]{\ifcase#1\or*\or$\dagger$\or$\ddagger$\or**\or$\dagger\dagger$\or$\ddagger\ddagger$\fi}
\makeatother
\author{H.E.S.S. Collaboration
\and H.~Abdalla \inst{\ref{NWU}}
\and F.~Aharonian \inst{\ref{MPIK},\ref{DIAS},\ref{RAU}}
\and F.~Ait~Benkhali \inst{\ref{MPIK}}
\and E.O.~Ang\"uner \inst{\ref{CPPM}}
\and M.~Arakawa \inst{\ref{Rikkyo}}
\and C.~Arcaro \inst{\ref{NWU}}
\and C.~Armand \inst{\ref{LAPP}}
\and H.~Ashkar \inst{\ref{IRFU}}
\and M.~Backes \inst{\ref{UNAM},\ref{NWU}}
\and V.~Barbosa~Martins \inst{\ref{DESY}}
\and M.~Barnard \inst{\ref{NWU}}
\and Y.~Becherini \inst{\ref{Linnaeus}}
\and D.~Berge \inst{\ref{DESY}}
\and K.~Bernl\"ohr \inst{\ref{MPIK}}
\and R.~Blackwell \inst{\ref{Adelaide}}
\and M.~B\"ottcher \inst{\ref{NWU}}
\and C.~Boisson \inst{\ref{LUTH}}
\and J.~Bolmont \inst{\ref{LPNHE}}
\and S.~Bonnefoy \inst{\ref{DESY}}
\and J.~Bregeon \inst{\ref{LUPM}}
\and M.~Breuhaus \inst{\ref{MPIK}}
\and F.~Brun \inst{\ref{IRFU}}
\and P.~Brun \inst{\ref{IRFU}}
\and M.~Bryan \inst{\ref{GRAPPA}}
\and M.~B\"{u}chele \inst{\ref{ECAP}}
\and T.~Bulik \inst{\ref{UWarsaw}}
\and T.~Bylund \inst{\ref{Linnaeus}}
\and M.~Capasso \inst{\ref{IAAT}}
\and S.~Caroff \inst{\ref{LPNHE}}
\and A.~Carosi \inst{\ref{LAPP}}
\and S.~Casanova \inst{\ref{IFJPAN},\ref{MPIK}}
\and M.~Cerruti \inst{\ref{LPNHE},\ref{CerrutiNowAt}}
\and N.~Chakraborty \inst{\ref{MPIK}}
\and T.~Chand \inst{\ref{NWU}}
\and S.~Chandra \inst{\ref{NWU}}
\and R.C.G.~Chaves \inst{\ref{LUPM},\ref{CurieChaves}}
\and A.~Chen \inst{\ref{WITS}}
\and S.~Colafrancesco \inst{\ref{WITS}} \protect\footnotemark[2] 
\and M.~Curylo \inst{\ref{UJK}}
\and I.D.~Davids \inst{\ref{UNAM}}
\and C.~Deil \inst{\ref{MPIK}}
\and J.~Devin \inst{\ref{CENBG}}
\and P.~deWilt \inst{\ref{Adelaide}}
\and L.~Dirson \inst{\ref{HH}}
\and A.~Djannati-Ata\"i \inst{\ref{APC}}
\and A.~Dmytriiev \inst{\ref{LUTH}}
\and A.~Donath \inst{\ref{MPIK}}
\and V.~Doroshenko \inst{\ref{IAAT}}
\and L.O'C.~Drury \inst{\ref{DIAS}}
\and J.~Dyks \inst{\ref{NCAC}}
\and K.~Egberts \inst{\ref{UP}}
\and G.~Emery \inst{\ref{LPNHE}}
\and J.-P.~Ernenwein \inst{\ref{CPPM}}
\and S.~Eschbach \inst{\ref{ECAP}}
\and K.~Feijen \inst{\ref{Adelaide}}
\and S.~Fegan \inst{\ref{LLR}}
\and A.~Fiasson \inst{\ref{LAPP}}
\and G.~Fontaine \inst{\ref{LLR}}
\and S.~Funk \inst{\ref{ECAP}}
\and M.~F\"u{\ss}ling \inst{\ref{DESY}}
\and S.~Gabici \inst{\ref{APC}}
\and Y.A.~Gallant \inst{\ref{LUPM}}
\and F.~Gat{\'e} \inst{\ref{LAPP}}
\and G.~Giavitto \inst{\ref{DESY}}
\and D.~Glawion \inst{\ref{LSW}}
\and J.F.~Glicenstein \inst{\ref{IRFU}}
\and D.~Gottschall \inst{\ref{IAAT}}
\and M.-H.~Grondin \inst{\ref{CENBG}}
\and J.~Hahn \inst{\ref{MPIK}}
\and M.~Haupt \inst{\ref{DESY}}
\and G.~Heinzelmann \inst{\ref{HH}}
\and G.~Henri \inst{\ref{Grenoble}}
\and G.~Hermann \inst{\ref{MPIK}}
\and J.A.~Hinton \inst{\ref{MPIK}}
\and W.~Hofmann \inst{\ref{MPIK}}
\and C.~Hoischen \inst{\ref{UP}}
\and T.~L.~Holch \inst{\ref{HUB}}
\and M.~Holler \inst{\ref{LFUI}}
\and D.~Horns \inst{\ref{HH}}
\and D.~Huber \inst{\ref{LFUI}}
\and H.~Iwasaki \inst{\ref{Rikkyo}}
\and M.~Jamrozy \inst{\ref{UJK}}
\and D.~Jankowsky \inst{\ref{ECAP}}
\and F.~Jankowsky \inst{\ref{LSW}}
\and I.~Jung-Richardt \inst{\ref{ECAP}}
\and M.A.~Kastendieck \inst{\ref{HH}}
\and K.~Katarzy{\'n}ski \inst{\ref{NCUT}}
\and M.~Katsuragawa \inst{\ref{KAVLI}}
\and U.~Katz \inst{\ref{ECAP}}
\and D.~Khangulyan \inst{\ref{Rikkyo}}
\and B.~Kh\'elifi \inst{\ref{APC}}
\and J.~King \inst{\ref{LSW}}
\and S.~Klepser \inst{\ref{DESY}}
\and W.~Klu\'{z}niak \inst{\ref{NCAC}}
\and Nu.~Komin \inst{\ref{WITS}}
\and K.~Kosack \inst{\ref{IRFU}}
\and D.~Kostunin \inst{\ref{DESY}} 
\and M.~Kraus \inst{\ref{ECAP}}
\and G.~Lamanna \inst{\ref{LAPP}}
\and J.~Lau \inst{\ref{Adelaide}}
\and A.~Lemi\`ere \inst{\ref{APC}}
\and M.~Lemoine-Goumard \inst{\ref{CENBG}}
\and J.-P.~Lenain \inst{\ref{LPNHE}}
\and E.~Leser \inst{\ref{UP},\ref{DESY}}
\and C.~Levy \inst{\ref{LPNHE}}
\and T.~Lohse \inst{\ref{HUB}}
\and R.~L\'opez-Coto \inst{\ref{MPIK}}
\and I.~Lypova \inst{\ref{DESY}}
\and J.~Mackey \inst{\ref{DIAS}}
\and J.~Majumdar \inst{\ref{DESY}}
\and D.~Malyshev \inst{\ref{IAAT}}
\and V.~Marandon \inst{\ref{MPIK}}
\and A.~Marcowith \inst{\ref{LUPM}}
\and A.~Mares \inst{\ref{CENBG}}
\and C.~Mariaud \inst{\ref{LLR}}
\and G.~Mart\'i-Devesa \inst{\ref{LFUI}}
\and R.~Marx \inst{\ref{MPIK}}
\and G.~Maurin \inst{\ref{LAPP}}
\and P.J.~Meintjes \inst{\ref{UFS}}
\and A.M.W.~Mitchell \inst{\ref{MPIK},\ref{MitchellNowAt}}
\and R.~Moderski \inst{\ref{NCAC}}
\and M.~Mohamed \inst{\ref{LSW}}
\and L.~Mohrmann \inst{\ref{ECAP}}
\and J.~Muller \inst{\ref{LLR}}
\and C.~Moore \inst{\ref{Leicester}}
\and E.~Moulin \inst{\ref{IRFU}}
\and T.~Murach \inst{\ref{DESY}}
\and S.~Nakashima  \inst{\ref{RIKKEN}}
\and M.~de~Naurois \inst{\ref{LLR}}
\and H.~Ndiyavala  \inst{\ref{NWU}}
\and F.~Niederwanger \inst{\ref{LFUI}}
\and J.~Niemiec \inst{\ref{IFJPAN}}
\and L.~Oakes \inst{\ref{HUB}}
\and P.~O'Brien \inst{\ref{Leicester}}
\and H.~Odaka \inst{\ref{Tokyo}}
\and S.~Ohm \inst{\ref{DESY}}
\and E.~de~Ona~Wilhelmi \inst{\ref{DESY}}
\and M.~Ostrowski \inst{\ref{UJK}}
\and I.~Oya \inst{\ref{DESY}}
\and M.~Panter \inst{\ref{MPIK}}
\and R.D.~Parsons \inst{\ref{MPIK}}
\and C.~Perennes \inst{\ref{LPNHE}}
\and P.-O.~Petrucci \inst{\ref{Grenoble}}
\and B.~Peyaud \inst{\ref{IRFU}}
\and Q.~Piel \inst{\ref{LAPP}}
\and S.~Pita \inst{\ref{APC}}
\and V.~Poireau \inst{\ref{LAPP}}
\and A.~Priyana~Noel \inst{\ref{UJK}}
\and D.A.~Prokhorov \inst{\ref{WITS}}
\and H.~Prokoph \inst{\ref{DESY}}
\and G.~P\"uhlhofer \inst{\ref{IAAT}}
\and M.~Punch \inst{\ref{APC},\ref{Linnaeus}}
\and A.~Quirrenbach \inst{\ref{LSW}}
\and S.~Raab \inst{\ref{ECAP}}
\and R.~Rauth \inst{\ref{LFUI}}
\and A.~Reimer \inst{\ref{LFUI}}
\and O.~Reimer \inst{\ref{LFUI}}
\and Q.~Remy \inst{\ref{LUPM}}
\and M.~Renaud \inst{\ref{LUPM}}\protect\footnotemark[1]
\and F.~Rieger \inst{\ref{MPIK}}
\and L.~Rinchiuso \inst{\ref{IRFU}}
\and C.~Romoli \inst{\ref{MPIK}}
\and G.~Rowell \inst{\ref{Adelaide}}
\and B.~Rudak \inst{\ref{NCAC}}
\and E.~Ruiz-Velasco \inst{\ref{MPIK}}
\and V.~Sahakian \inst{\ref{YPI}}
\and S.~Saito \inst{\ref{Rikkyo}}
\and D.A.~Sanchez \inst{\ref{LAPP}}
\and A.~Santangelo \inst{\ref{IAAT}}
\and M.~Sasaki \inst{\ref{ECAP}}
\and R.~Schlickeiser \inst{\ref{RUB}}
\and F.~Sch\"ussler \inst{\ref{IRFU}}
\and A.~Schulz \inst{\ref{DESY}}
\and H.~Schutte \inst{\ref{NWU}}
\and U.~Schwanke \inst{\ref{HUB}}
\and S.~Schwemmer \inst{\ref{LSW}}
\and M.~Seglar-Arroyo \inst{\ref{IRFU}}
\and M.~Senniappan \inst{\ref{Linnaeus}}
\and A.S.~Seyffert \inst{\ref{NWU}}
\and N.~Shafi \inst{\ref{WITS}}
\and K.~Shiningayamwe \inst{\ref{UNAM}}
\and R.~Simoni \inst{\ref{GRAPPA}}\protect\footnotemark[1]
\and A.~Sinha \inst{\ref{APC}}
\and H.~Sol \inst{\ref{LUTH}}
\and A.~Specovius \inst{\ref{ECAP}}
\and M.~Spir-Jacob \inst{\ref{APC}}
\and {\L.}~Stawarz \inst{\ref{UJK}}
\and R.~Steenkamp \inst{\ref{UNAM}}
\and C.~Stegmann \inst{\ref{UP},\ref{DESY}}
\and C.~Steppa \inst{\ref{UP}}
\and T.~Takahashi  \inst{\ref{KAVLI}}
\and T.~Tavernier \inst{\ref{IRFU}}
\and A.M.~Taylor \inst{\ref{DESY}}
\and R.~Terrier \inst{\ref{APC}}
\and D.~Tiziani \inst{\ref{ECAP}}
\and M.~Tluczykont \inst{\ref{HH}}
\and C.~Trichard \inst{\ref{LLR}}
\and M.~Tsirou \inst{\ref{LUPM}}
\and N.~Tsuji \inst{\ref{Rikkyo}}
\and R.~Tuffs \inst{\ref{MPIK}}
\and Y.~Uchiyama \inst{\ref{Rikkyo}}
\and D.J.~van~der~Walt \inst{\ref{NWU}}
\and C.~van~Eldik \inst{\ref{ECAP}}
\and C.~van~Rensburg \inst{\ref{NWU}}
\and B.~van~Soelen \inst{\ref{UFS}}
\and G.~Vasileiadis \inst{\ref{LUPM}}
\and J.~Veh \inst{\ref{ECAP}}
\and C.~Venter \inst{\ref{NWU}}
\and P.~Vincent \inst{\ref{LPNHE}}
\and J.~Vink \inst{\ref{GRAPPA}}\protect\footnotemark[1]
\and F.~Voisin \inst{\ref{Adelaide}}
\and H.J.~V\"olk \inst{\ref{MPIK}}
\and T.~Vuillaume \inst{\ref{LAPP}}
\and Z.~Wadiasingh \inst{\ref{NWU}}
\and S.J.~Wagner \inst{\ref{LSW}}
\and R.~White \inst{\ref{MPIK}}
\and A.~Wierzcholska \inst{\ref{IFJPAN},\ref{LSW}}
\and R.~Yang \inst{\ref{MPIK}}
\and H.~Yoneda \inst{\ref{KAVLI}}
\and M.~Zacharias \inst{\ref{NWU}}
\and R.~Zanin \inst{\ref{MPIK}}
\and A.A.~Zdziarski \inst{\ref{NCAC}}
\and A.~Zech \inst{\ref{LUTH}}
\and A.~Ziegler \inst{\ref{ECAP}}
\and J.~Zorn \inst{\ref{MPIK}}
\and N.~\.Zywucka \inst{\ref{NWU}}
\and N.I.~Maxted \inst{\ref{UNSWCAN}}\protect\footnotemark[1]
}

\institute{
Centre for Space Research, North-West University, Potchefstroom 2520, South Africa \label{NWU} \and 
Universit\"at Hamburg, Institut f\"ur Experimentalphysik, Luruper Chaussee 149, D 22761 Hamburg, Germany \label{HH} \and 
Max-Planck-Institut f\"ur Kernphysik, P.O. Box 103980, D 69029 Heidelberg, Germany \label{MPIK} \and 
Dublin Institute for Advanced Studies, 31 Fitzwilliam Place, Dublin 2, Ireland \label{DIAS} \and 
High Energy Astrophysics Laboratory, RAU,  123 Hovsep Emin St  Yerevan 0051, Armenia \label{RAU} \and
Yerevan Physics Institute, 2 Alikhanian Brothers St., 375036 Yerevan, Armenia \label{YPI} \and
Institut f\"ur Physik, Humboldt-Universit\"at zu Berlin, Newtonstr. 15, D 12489 Berlin, Germany \label{HUB} \and
University of Namibia, Department of Physics, Private Bag 13301, Windhoek, Namibia, 12010 \label{UNAM} \and
GRAPPA, Anton Pannekoek Institute for Astronomy, University of Amsterdam,  Science Park 904, 1098 XH Amsterdam, The Netherlands \label{GRAPPA} \and
Department of Physics and Electrical Engineering, Linnaeus University,  351 95 V\"axj\"o, Sweden \label{Linnaeus} \and
Institut f\"ur Theoretische Physik, Lehrstuhl IV: Weltraum und Astrophysik, Ruhr-Universit\"at Bochum, D 44780 Bochum, Germany \label{RUB} \and
Institut f\"ur Astro- und Teilchenphysik, Leopold-Franzens-Universit\"at Innsbruck, A-6020 Innsbruck, Austria \label{LFUI} \and
School of Physical Sciences, University of Adelaide, Adelaide 5005, Australia \label{Adelaide} \and
LUTH, Observatoire de Paris, PSL Research University, CNRS, Universit\'e Paris Diderot, 5 Place Jules Janssen, 92190 Meudon, France \label{LUTH} \and
Sorbonne Universit\'e, Universit\'e Paris Diderot, Sorbonne Paris Cit\'e, CNRS/IN2P3, Laboratoire de Physique Nucl\'eaire et de Hautes Energies, LPNHE, 4 Place Jussieu, F-75252 Paris, France \label{LPNHE} \and
Laboratoire Univers et Particules de Montpellier, Universit\'e Montpellier, CNRS/IN2P3,  CC 72, Place Eug\`ene Bataillon, F-34095 Montpellier Cedex 5, France \label{LUPM} \and
IRFU, CEA, Universit\'e Paris-Saclay, F-91191 Gif-sur-Yvette, France \label{IRFU} \and
Astronomical Observatory, The University of Warsaw, Al. Ujazdowskie 4, 00-478 Warsaw, Poland \label{UWarsaw} \and
Aix Marseille Universit\'e, CNRS/IN2P3, CPPM, Marseille, France \label{CPPM} \and
Instytut Fizyki J\c{a}drowej PAN, ul. Radzikowskiego 152, 31-342 Krak{\'o}w, Poland \label{IFJPAN} \and
Funded by EU FP7 Marie Curie, grant agreement No. PIEF-GA-2012-332350 \label{CurieChaves}  \and
School of Physics, University of the Witwatersrand, 1 Jan Smuts Avenue, Braamfontein, Johannesburg, 2050 South Africa \label{WITS} \and
Laboratoire d'Annecy de Physique des Particules, Univ. Grenoble Alpes, Univ. Savoie Mont Blanc, CNRS, LAPP, 74000 Annecy, France \label{LAPP} \and
Landessternwarte, Universit\"at Heidelberg, K\"onigstuhl, D 69117 Heidelberg, Germany \label{LSW} \and
Universit\'e Bordeaux, CNRS/IN2P3, Centre d'\'Etudes Nucl\'eaires de Bordeaux Gradignan, 33175 Gradignan, France \label{CENBG} \and
Oskar Klein Centre, Department of Physics, Stockholm University, Albanova University Center, SE-10691 Stockholm, Sweden \label{OKC} \and
Institut f\"ur Astronomie und Astrophysik, Universit\"at T\"ubingen, Sand 1, D 72076 T\"ubingen, Germany \label{IAAT} \and
Laboratoire Leprince-Ringuet, École Polytechnique, UMR 7638, CNRS/IN2P3, Institut Polytechnique de Paris, F-91128 Palaiseau, France \label{LLR} \and
APC, AstroParticule et Cosmologie, Universit\'{e} Paris Diderot, CNRS/IN2P3, CEA/Irfu, Observatoire de Paris, Sorbonne Paris Cit\'{e}, 10, rue Alice Domon et L\'{e}onie Duquet, 75205 Paris Cedex 13, France \label{APC} \and
Univ. Grenoble Alpes, CNRS, IPAG, F-38000 Grenoble, France \label{Grenoble} \and
Department of Physics and Astronomy, The University of Leicester, University Road, Leicester, LE1 7RH, United Kingdom \label{Leicester} \and
Nicolaus Copernicus Astronomical Center, Polish Academy of Sciences, ul. Bartycka 18, 00-716 Warsaw, Poland \label{NCAC} \and
Institut f\"ur Physik und Astronomie, Universit\"at Potsdam,  Karl-Liebknecht-Strasse 24/25, D 14476 Potsdam, Germany \label{UP} \and
Friedrich-Alexander-Universit\"at Erlangen-N\"urnberg, Erlangen Centre for Astroparticle Physics, Erwin-Rommel-Str. 1, D 91058 Erlangen, Germany \label{ECAP} \and
DESY, D-15738 Zeuthen, Germany \label{DESY} \and
Obserwatorium Astronomiczne, Uniwersytet Jagiello{\'n}ski, ul. Orla 171, 30-244 Krak{\'o}w, Poland \label{UJK} \and
Centre for Astronomy, Faculty of Physics, Astronomy and Informatics, Nicolaus Copernicus University,  Grudziadzka 5, 87-100 Torun, Poland \label{NCUT} \and
Department of Physics, University of the Free State,  PO Box 339, Bloemfontein 9300, South Africa \label{UFS} \and
Department of Physics, Rikkyo University, 3-34-1 Nishi-Ikebukuro, Toshima-ku, Tokyo 171-8501, Japan \label{Rikkyo} \and
Kavli Institute for the Physics and Mathematics of the Universe (WPI), The University of Tokyo Institutes for Advanced Study (UTIAS), The University of Tokyo, 5-1-5 Kashiwa-no-Ha, Kashiwa City, Chiba, 277-8583, Japan \label{KAVLI} \and
Department of Physics, The University of Tokyo, 7-3-1 Hongo, Bunkyo-ku, Tokyo 113-0033, Japan \label{Tokyo} \and
RIKEN, 2-1 Hirosawa, Wako, Saitama 351-0198, Japan \label{RIKKEN} \and
Now at Physik Institut, Universit\"at Z\"urich, Winterthurerstrasse 190, CH-8057 Z\"urich, Switzerland \label{MitchellNowAt} \and
Now at Institut de Ci\`{e}ncies del Cosmos (ICC UB), Universitat de Barcelona (IEEC-UB), Mart\'{i} Franqu\`es 1, E08028 Barcelona, Spain \label{CerrutiNowAt} \and School of Sciences, University of New South Wales, Australian Defence Force Academy, Canberra, ACT 2600, Australia \label{UNSWCAN}
}
 
\offprints{H.E.S.S.~collaboration,
\protect\\\email{\href{mailto:contact.hess@hess-experiment.eu}{contact.hess@hess-experiment.eu}};
\protect\\${}^*$ Corresponding author
\protect\\${}^\dagger$ Deceased}



 \date{Received 10/02/2019; Accepted 15/04/2019}

  \abstract
   {Young core-collapse supernovae with dense-wind progenitors may be able to accelerate cosmic-ray hadrons beyond the knee of the {cosmic-ray spectrum}, and this may result in measurable gamma-ray emission.We searched for gamma-ray emission from ten supernovae observed with the High Energy Stereoscopic System (H.E.S.S.) within a year of the supernova event. Nine supernovae were observed serendipitously in the H.E.S.S. data collected between December 2003 and December 2014, with exposure times ranging from 1.4 hours to 53 hours. In addition we observed SN 2016adj as a target of opportunity in February 2016 for 13 hours. No significant gamma-ray emission {has been} detected for any of the objects, and upper limits on the $>1$~TeV gamma-ray flux of the order of $\sim$10$^{-13}$\,cm$^{-2}$s$^{-1}$ {are} established, corresponding to upper limits on the luminosities in the range $\sim$2\,$\times$\,10$^{39}$\, erg\,s $^{-1}$ {to} $\sim$1\,$\times$\,10$^{42}$\, erg\,s$^{-1}$. These values are used to place model-dependent constraints on the mass-loss rates of the progenitor stars, implying upper limits {between $\sim$2 $\times\,10^{-5}$ and} $\sim$2 $\times\,10^{-3}$\,M$_{\odot}$\,yr$^{-1}$ under reasonable assumptions on the particle acceleration parameters.}


\maketitle
\makeatletter
\renewcommand*{\@fnsymbol}[1]{\ifcase#1\@arabic{#1}\fi}
\makeatother
\section{Introduction}
\label{sec:Intro}
Despite more than a century of cosmic-ray (CR) studies, the origin of {CR}s\footnote{In the manuscript, CR refers to cosmic-ray hadrons, nuclei and electrons when not otherwise specified.} is still not clear. The bulk of {CRs} detected  {at} Earth or in space are of Galactic origin, at least for protons up to energies of $3\times 10^{15}$~eV \citep[e.g.][]{Ginzburg:1964,Strong:2007}. Galactic CR sources can only be identified indirectly through the electromagnetic and neutrino emission caused by interactions of {CRs} with local gas and radiation within their sources of origin. In this respect, gamma-ray observations have proven to be an invaluable tool, as gamma-ray emission is uniquely associated with the presence of highly energetic {CR particles}{. Gamma rays} can be detected with sufficient statistics using gamma-ray observatories, either on-board satellites such as NASA's {\it Fermi} Large Area Telescope (LAT) {in the high-energy gamma-ray domain (HE, 0.1\,$<\mathrm{E} <100$~GeV), or ground-based 
Cherenkov telescopes, such as H.E.S.S., {VERITAS}, MAGIC {and HAWC} in the very high-energy gamma-ray domain (VHE, i.e. $\mathrm{E} \gtrsim 50$~GeV)}.

These instruments show that the prime candidate sources for Galactic CRs, supernova remnants {(SNRs)}, are indeed gamma-ray sources.{ Young SNRs, i.e. those SNRs that are a few hundred to a few thousand years of age}, emit TeV gamma rays at various stages during their evolution \citep[][]{Hewitt:2015}. However, there is an ongoing debate {as to} whether the emission is the result of hadronic CRs interacting with ambient gas, or leptonic interactions, usually inverse Compton scattering of ambient photons by relativistic electrons. For some older remnants interacting with dense gas, the gamma-ray emission is clearly associated with hadronic CRs: 
{the gamma-ray spectrum {below GeV energies shows the predicted spectral feature of hadronic gamma-ray emission, usually referred to as the "pion bump"} \citep[][]{Giuliani:2011,Ackermann:2013}.} However, there is currently no observational evidence that SNRs contain CR hadrons with energies around 10$^{15}$~eV (=1 PeV) or {above, even for the youngest Galactic SNRs such as Cas A \citep{Ahnen:2017}. {This is somewhat at odds} with the fact that Galactic CR hadrons need to be accelerated up to energies of at least $3\times 10^{15}$~eV to explain the knee of the CR spectrum.} 

The energy per supernova {(SN)} coupled with the Galactic {SN} rate provides enough power to sustain the CR energy density in the Galaxy \citep{Strong:2004}. 
Moreover, it has been suggested that, for supernova remnants evolving in the winds of their progenitors, the maximum CR energy is reached in the early phase of the SNR development, { within days to months after the SN event} and not at the time the {SNR} is several hundred
to thousand years old \citep[e.g.][]{Voelk:1988,Cardillo:2015, Marcowith:2018}. The highest energy CR hadrons are then {possibly} accelerated in this early phase, at least for a subset of {SNe} {\citep{Bell:2013,Zirakashvili:2016}}.
At this early stage, less time is available for acceleration, and the shock surface area is much smaller.  
{Only SNe exploding into sufficiently large circumstellar densities }
can accelerate sufficient numbers of CR proton up to energies of, or exceeding PeV energies: {the number of CR hadrons being accelerated at a given time depends on the local density $n$ by the relation $\dot{N}_\mathrm{CR}\propto nR_\mathrm{sh}^2 V_\mathrm{sh}$,} where $R_\mathrm{sh}$ is the shock radius and $V_\mathrm{sh}$ is the shock velocity. 
{Furthermore, the maximum energy that can be reached at given time is dependant on the magnetic field and the turbulence as shown by the following relation \citep[e.g.][]{Helder:2012} : 
\begin{equation}
\label{eq:tacc}
\begin{split}
t = &~\eta_\mathrm{acc}\eta_\mathrm{g} \frac{cE}{3eBV_\mathrm{sh}^2}  \\
\nonumber\approx  &1.0\times 10^4 \left(\frac{\eta_\mathrm{g}\eta_\mathrm{acc}}{30}\right)\left(\frac{V_\mathrm{sh}}{10,000~\mathrm{km\,s^{-1}}}\right)^{-2}\left(\frac{B}{10~\mathrm{G}}\right)^{-1} \left(\frac{E}{100~\mathrm{TeV}}\right)~,
\end{split}
\end{equation}
with $V_\mathrm{sh}$ the shock velocity, $B$ is the magnetic field strength, $t$ is the time available for acceleration in seconds, $\eta_\mathrm{g}$ indicates the ratio of mean free path of the particles 
and their gyroradius ($\eta_\mathrm{g}\approx 1$, for a very turbulent magnetic field) and $\eta_\mathrm{acc}=8-20$ takes into account the difference in occupancy time of particles upstream
and downstream of the shock. 
This equation shows that in order to reach energies of 100 TeV, typically an acceleration time shorter than a day is sufficient, and for reaching the CR knee at $3\times 10^{15}$~eV,
an acceleration time of days to weeks is needed 
provided that the magnetic fields are $>1$~G, instead of the $10-100~\mu$G  measured in young SNRs. 
More detailed calculations that also take into account the expected evolution of the magnetic field strength and escape of the highest energy particles
confirm these time scales, see 
\cite{Tatischeff:2009,Marcowith:2018}. 
}

{Core-collapse (cc-)SNe} originating from {stellar} progenitors with dense winds can fulfil the {right} conditions for CR acceleration, provided that the shocks are collisionless \citep{Katz:2011,Murase:2011,Bell:2013}. First of all, the number density of a stellar wind scales as a function of radius, $n \propto \dot{M}u_\mathrm{w}^{-1}r^{-2}$ (where $\dot{M}$ is the mass-loss rate and $u_\mathrm{w}$ is the wind velocity), resulting in a circumstellar medium (CSM) density that is much higher than in the interstellar medium (ISM).
{As a result the  cosmic-ray acceleration rate does not directly depend on the shock radius, but scales as $\dot{N}_\mathrm{CR} \propto \dot{M}u_\mathrm{w}^{-1}V_\mathrm{sh}$.}
{Secondly, the magnetic fields around the shock can be amplified early on to values orders of magnitude larger than in the {ISM}, by the growth of CR streaming instabilities \citep{Bell:2004}. {Once CR acceleration begins}, {it has been argued that magnetic field strengths may be amplified scaling as $B^2 \propto n V_\mathrm{sh}^2$} \citep[e.g. see][]{Voelk:2005,Bell:2013} or even $B^2 \propto n V_\mathrm{sh}^3$ \citep[][]{Vink:review,Bell:2004}, with shock velocities as fast as 20,000 km s$^{-1}$}.  {SNe arising} from progenitors that exhibit these high-density stellar winds are of type IIP, IIL, IIb and IIn \citep[see the review in][]{Chevalier:review}. Type IIn SNe, in particular, have been observed to eject a considerable mass prior to the explosion \citep{Smith:2008,Ofek:2014,Murase:2014}.

Evidence for particle acceleration in {cc-}SNe {is} provided by their bright, self-absorbed synchrotron emission at radio wavelengths. These radio observations indicate the presence of relativistic electrons accelerated by strong magnetic fields, and it is thought that relativistic protons and atomic nuclei are present. One prime example is the young type IIb SN\,1993J, for which a strong magnetic field, 1-100 G, { and shock speeds as high as 20,000 km\,s$^{-1}$ have} been {estimated} by \citet{Fransson:1998} and \citet{Tatischeff:2009}. 
{Given these estimates and  time scales for acceleration to PeV energies, there are good reasons to think that SN 1993J-like SNe can accelerate CR hadrons up to PeV energies { within days to a few weeks} \citep[see][and references therein]{Marcowith:2018}. {However,  gamma-ray emission from SNe has not been detected so far}, and upper-limits {in the GeV domain} have been established for type IIn {SNe} \citep{Ackermann:2015}, and superluminous {SN} candidates \citep{Renault-Tinacci:2017}.
The possibility of CR acceleration in young SNe motivates the search for signatures of TeV gamma-ray emission. {Such a detection {could} be indicative of pion production {(and subsequent decay) arising from} acceleration of CR protons and nuclei beyond TeV energies. We analysed H.E.S.S. {observations} towards {nine} serendipitously observed {cc-}SN events and {towards the nearby event} SN 2016adj, which triggered Target of Opportunity (ToO) observations. The data selection and analysis details are presented in Section\,\ref{sec:Candidates}. Upper limits on gamma-ray {flux from the direction of} the CR source candidates are shown in Section\,\ref{sec:Results} before constraints on {the local environment of} these SNe are derived in Section\,\ref{sec:discuss}.
\section{Observations and data analysis}\label{sec:obs}

\subsection{H.E.S.S. observations}

H.E.S.S., the High Energy Stereoscopic System, is an array of five imaging atmospheric Cherenkov telescopes (IACTs) located in the Khomas Highland of Namibia at an altitude of 1800\,m above sea level. Four 12\,m-diameter telescopes {(CT1-4)} {have been} operating from December 2003 and a {fifth} telescope of 28\,m-diameter {(CT5)} became operational in September 2012. In the analysis, the data of {at least {three} telescopes (including CT5 for SN 2016adj)} have been utilised. {These analysis settings correspond to} a field of view of 5$^{\circ}$, an angular resolution (68$\%$ containment radius) of $\sim$0.1$^{\circ}$, energy threshold values spanning $\sim$200\,GeV {to} $\sim$400\,GeV and an energy resolution of $\sim$15$\%$, \citep{Aharonian:2006}. Nine SNe in our sample were in the field of view of other H.E.S.S.-scheduled targets, with a maximal offset of $2.5^{\circ}$ {from the center of the field of view} : the candidate selection is described in Section\,\ref{sec:serendipitous}. {The nearby event SN 2016adj {(see Section \ref{sec:SN2016adj})} triggered dedicated observations a} few days after the discovery date, in standard wobble mode observations \citep{Aharonian:2006} with {a source offset of $\sim$0.5$^{\circ}$}.

\subsection{Supernova Selection}\label{sec:Candidates}

\subsubsection{Serendipitous sample}\label{sec:serendipitous}

The online IAU\footnote{International Astronomical Union} Central Bureau of Astronomical Telegrams (CBAT) supernova catalogue\footnote{www.cbat.eps.harvard.edu/lists/Supernovae.html} was used  to compile an initial, extensive list of SN candidates. The NASA/IPAC\footnote{Infrared Processing $\&$ Analysis Center} Extragalactic Database (NED)\footnote{ned.ipac.caltech.edu/forms/z.html} was then queried for the redshift of each SN host galaxy to compile a short-list of SNe with redshift $z<$\,0.01 to ensure that only nearby SNe were considered. If a host galaxy was not stated for a given SN in the CBAT SN catalogue, the SN was discarded from the short-list. The H.E.S.S. database was then queried for observations in the direction of each short-listed SN, within a {time range spanning seven days prior to a year after,} the SN discovery date. This time range was chosen to account for likely delays between the dates of the discovery and the outburst {to be sure to include} the peak energy suggested to occur {a} few days \citep{Marcowith:2014} after the SN event, but {potentially lasting} months. After reaching this peak emission, the gamma-ray flux is {predicted} to decline proportionally to $1/t$, as we will discuss later. All H.E.S.S. data taken from December 2003 until the 31st of December 2014 {were} searched, and all SNe presented on CBAT on the 30th of March 2015 were considered. We removed type Ia and Ic SNe from our sample, because these types are unlikely to occur in a CSM density large enough to accelerate CRs up to TeV energies \citep[see e.g.][]{Smith:2014}.

\begin{table*}
\caption{The list of SN positions tested for H.E.S.S. gamma-ray excess emission. The list was compiled using a system of cuts described in Section\,\ref{sec:Candidates}. The name, host galaxy, coordinates, estimated distance, SN type and discovery date {are given} for each SN. \label{tab:SNcrit}}
\centering
\begin{tabular}{|l|l|l|l|l|l|l|}
\hline
\textbf{SN Name} & \textbf{Host galaxy} 		& \textbf{RA [J2000]} & \textbf{DEC [J2000]} & \textbf{Dist. [Mpc]} & \textbf{Type} & \textbf{Disc. date}\TBstrut\\
\hline			
SN\,2004cx		& NGC\,7755	& 23h47m52.86s 	& $-$30$^{\circ}$31$^{\prime}$32.6$^{\prime\prime}$	&	26\,$\pm$\,5 	& II 	& 2004-06-26\Tstrut \\
SN\,2005dn		& NGC\,6861	& 20h11m11.73s 	& $-$48$^{\circ}$16$^{\prime}$35.5$^{\prime\prime}$	&	38.4\,$\pm$\,2.7	& II& 2005-08-27	 \\
SN\,2008bk		& NGC\,7793	& 23h57m50.42s 	& $-$32$^{\circ}$33$^{\prime}$21.5$^{\prime\prime}$	&	4.0\,$\pm$\,0.4 	& IIP 	& 2008-03-25	 \\
SN\,2008bp		& NGC\,3095	& 10h00m01.57s	& $-$31$^{\circ}$33$^{\prime}$21.8$^{\prime\prime}$	&	29\,$\pm$\,6 	& IIP 	& 2008-04-02	 \\
SN\,2008ho		& NGC\,922	& 02h25m04.00s 	& $-$24$^{\circ}$48$^{\prime}$02.4$^{\prime\prime}$	&	41.5\,$\pm$\,2.9& IIP 	& 2008-11-26 \\
SN\,2009hf		& NGC\,175	& 00h37m21.79s	& $-$19$^{\circ}$56$^{\prime}$42.2$^{\prime\prime}$	&	53.9\,$\pm$\,3.8	& IIP 	& 2009-07-09	 \\
SN\,2009js		& NGC\,918	& 02h25m48.28s 	& $+$18$^{\circ}$29$^{\prime}$25.8 $^{\prime\prime}$	&	16\,$\pm$\,3 	& IIP	& 2009-10-11	 \\
SN\,2011ja		& NGC\,4945	& 13h05m11.12s	& $-$49$^{\circ}$31$^{\prime}$27.0$^{\prime\prime}$	&	5.28\,$\pm$\,0.38&IIP	& 2011-12-18 \\
SN\,2012cc		& NGC\,4419	& 12h26m56.81s 	& $+$15$^{\circ}$02$^{\prime}$45.5$^{\prime\prime}$	&	16.5\,$\pm$1.1\,&II		& 2012-04-29 \\
SN\,2016adj     & NGC\,5128 & {13h25m24.11s} & {$-$43$^{\circ}$00$^{\prime}$57.5$^{\prime\prime}$} & 3.8\,$\pm${0.1}&IIb&2016-02-08\\
\hline
\end{tabular}
\end{table*}

\begin{figure*}
\begin{center}
\includegraphics[trim=1cm 1.2cm 0cm 0cm, width=0.9\textwidth,]{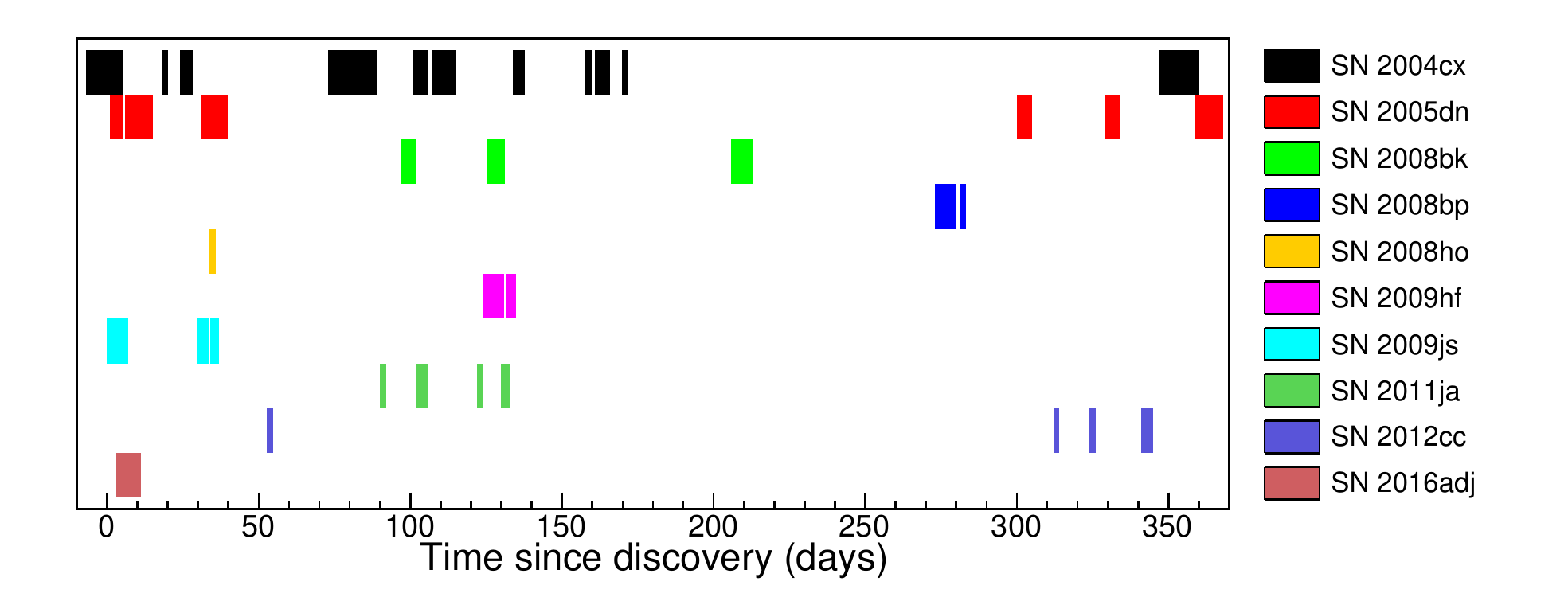}
\caption{A graphic displaying the H.E.S.S. observation windows for each SN {selected} for analysis. {Time $=0$ corresponds to the SN discovery date reported in literature.}} 
\label{fig:Dataset}
\end{center}
\end{figure*}

\subsubsection{SN 2016adj}\label{sec:SN2016adj}

SN 2016 adj was discovered on the 8th of February in the BOSS Survey\footnote{http://bosssupernova.com/}  \citep{ATel0802}. It appeared in the {Centaurus A (Cen A, NGC 5128) galaxy}, a few arcminutes from {its nucleus}, at a distance of {$\sim$3.8 $\pm$ 0.1~Mpc \cite[][and references therein]{Harris:2010}}. The SN type was {soon} suggested to be of type IIb by \citet{ATel902}, because of the presence of {a} H$\alpha$ line, and later confirmed by \citet{ATel1002}. {The proximity and the type of the SN made it a unique and rare event for triggering VHE gamma-ray observations: the last type IIb SN {occurring at a similar distance} was SN 1993J, and such a nearby cc-SN could be detected at TeV gamma rays with current IACTs as predicted by} \cite{Marcowith:2014}. Cen A {itself} was detected by H.E.S.S. after a deep observation campaign \citep{CenA:2009}. Cen A is quite faint at TeV energies and the exposure time of a few hours considered here for SN 2016adj is too short to detect emission from the galaxy itself. \\
{Table\,\ref{tab:SNcrit} lists the ten observed SNe along with their corresponding host galaxies, coordinates, distances {(ranging between $\sim$4 and 54\,Mpc)}, types and discovery dates.}

\subsection{Data Analysis}\label{sec:analysis}

Standard quality cuts were applied to remove bad-quality data from each data set: observation runs with a minimum of {three} telescopes, a fraction of broken pixels <\,10 $\%$ and  trigger-rate {fluctuations} <\,30 $\%$ were kept \citep[see][]{Aharonian:2006}. Then, for each target, the selected data were analysed using the Model analysis framework outlined in \citet{deNaurois:2009}. For a given SN, an ON-region is selected centred on the SN position, and multiple OFF regions are selected at the same offset as the ON region, using the Reflected Background method \citep{Berge:2007}. {The ON region radius is {0.1 deg} for the serendipitous sample and 0.08 deg for SN 2016adj.} Results were confirmed by an independent calibration and analysis chain using the ImPACT framework \citep{ImPACT} with standard quality cuts and the same Reflected Background method. For each 28-minute observation run which passed the above-mentioned criteria, the {gamma-ray} excess was computed using N$_\mathrm{excess}$ = N$_\mathrm{on} -\alpha$N$_\mathrm{off}$, with $\alpha$ being {the ratio of the solid angles of on and off regions.} The statistical significance for each dataset was established using equation\,17 of \citet{Li:1983}. 

\subsection{Results}\label{sec:Results}

No significant excess is observed for any of the SNe and flux upper limits (ULs) have been derived {at the 95$\%$ confidence level under the assumption of a {power law spectrum (dN/dE $\propto$ E$^{-\Gamma}$) with index $\Gamma=2$}.} {ULs have been computed using a loglikelihood approach as described in \cite{denaurois:tel-00687872}.}

In Table\,\ref{tab:statistics} we report the relevant statistics of the gamma-ray observations as described in Section\,\ref{sec:analysis}. The total livetime, and the observational {coverages} expressed in days since the SN discovery date are also presented. Given the serendipitous nature of {most of} the observations, the livetime varies between $\sim$1 and $\sim$50\,hours, and the time delay of the first observation {since the SN discovery} differs between {the }SNe: for {four objects (including SN 2016adj)}, {H.E.S.S. observations started around, or a few days after}, the discovery date, {while for the other SNe,} observations began much later (up to 272 days after the discovery for SN 2008bp). The observational {coverage} in days after the discovery date are reported in Table \,\ref{tab:statistics}  and the data set for each SN is represented in Figure \ref{fig:Dataset}.{The average time delays, weighted by the exposure of individual observation periods (as represented on Figure \ref{fig:Dataset}), are also} reported in Table \,\ref{tab:statistics}. SN 2016adj {was} observed every night from day 3 till day 10 after the discovery date, {and the average time delay amounts to $\sim$7\,days.}\\

In Table \ref{tab:ul}, {ULs on the integrated flux above the energy threshold and above 1 TeV are presented}. {The value of 1 TeV corresponds to the optimal H.E.S.S. sensitivity: it is chosen to compare all results, as the energy threshold depends on observational conditions and varies as indicated in the table.} For four SNe (2004cx, 2008bk, 2008bp, 2009js) these ULs supersede previous preliminary results \citep{Lennarz:2013}, confirming the non-detections with better sensitivity. In column five of Table \ref{tab:ul}, ULs on the luminosities are presented for each object: these ULs are computed above the energy threshold {and above 1 TeV}, using the distance to the host galaxy (see Table \ref{tab:SNcrit}). Errors on the distances are not taken into account.
{The {luminosity values {above the energy threshold} span} a range from $\sim$2 $\times$10$^{39}$\, erg\,s $^{-1}$ to $\sim$1 $\times$ 10$^{42}$\, erg\,s$^{-1}$. This range is mainly due to the differences in the source distances and to the offset angle with respect to the {observation position}, {observations with {large} off-center angles having a reduced detection sensitivity. 
Note that these ULs correspond to a gamma-ray fluence within a year of  $\sim$6 $\times$ 10$^{46}$\, erg {to} $\sim$3 $\times$ 10$^{49}$ erg, corresponding to 0.006 {-} 3 percent of the canonical SN explosion energy of 10$^{51}$ erg.}
We carefully checked that no significant gamma-ray peak occurred during the duration of the observations. As an example, Figure \ref{LC_2016adj} shows {the time evolution of the flux} above 1\,TeV  for SN 2016adj, consistent with zero during the {observing period}. This is the case for all {the other} SNe (see Appendix \ref{fig:LightCurves_all}). {The lightcurves presented are binned on a nightly basis and we also checked that no significant emission occurred on a weekly basis for any of the objects. Error bars correspond to 68$\%$ confidence levels.} This confirms {that} no significant TeV emission is found towards any of the SNe within one year of the initial explosion.

\begin{figure}
\begin{center}
\includegraphics[width=0.99\textwidth]{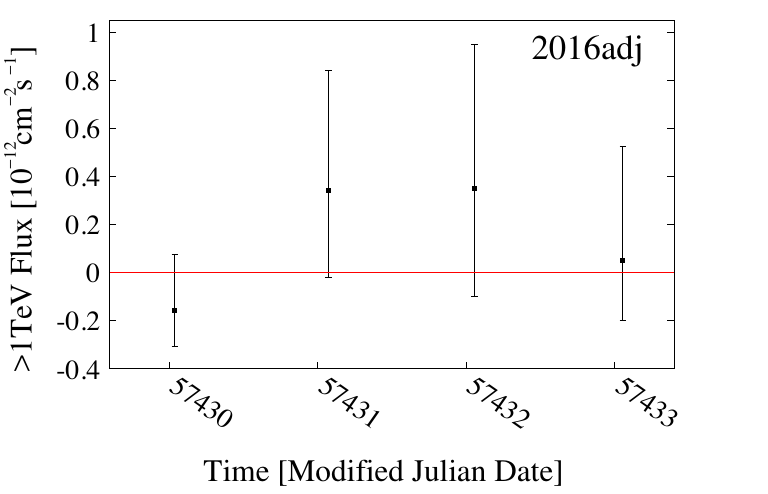}
\caption{ Light curve of SN 2016adj. {Data points are binned on a nightly basis}, {the red solid line indicates the zero level.}}
\label{LC_2016adj}
\end{center}
\end{figure}

\begin{table*}
\begin{tabular}{*{1}{|l} *{8}{|c} |}
    \hline
SNe& E$_\mathrm{Th}$ 			&	{\small Flux  UL}			&	{\small Flux UL} &\small{ UL on L}&\small{ UL on L}&\multicolumn{2}{c|}{\small{ULs on~$\dot{M }/{ u_\mathrm{w}}$}}  \Tstrut\\
&&\scriptsize{$ (>E_\mathrm{Th})$}&\scriptsize{$(>1\textrm{TeV})$}&\scriptsize{$ (>E_\mathrm{Th})$}&\scriptsize{ $(>1\textrm{TeV})$}&\tiny{Average time}& \tiny{Fit}\\
 
&   \scriptsize{(TeV)}	&	\multicolumn{2}{c|}{\scriptsize{(10$^{-13}$\,cm$^{-2}$s$^{-1}$)} }&\multicolumn{2}{c|}{\scriptsize{(10$^{40}$ erg\,s$^{-1}$) }}&\multicolumn{2}{c|}{\scriptsize{(10$^{-5}~\mathrm{M_{\odot}~yr^{-1}~km^{-1}\,s}$)}}\\
 \hline
 \small{SN\,2004cx}&    	0.18				&	10							&	1.9&	13.0&2.5&	6.7&	{3.2}\Tstrut\\
 \small{SN\,2005dn}&    	0.21				&	2.2						&	0.41&		6.2&1.2&	3.8&{0.26}\\
 \small{SN\,2008bk}&   0.21				&	6.0					&	4.8&		0.18	&0.15&	1.4&	{0.4}\\
 \small{SN\,2008bp}&    	0.21				&	29							&	5.5&		46.7&8.9&	15.9&{12.3}\\
 \small{SN\,2008ho}&   0.33				&	16							&	7.7&		52.8&25.4 &9.4	&{ 5.3}\\
 \small{SN\,2009hf}&  0.21				&	20						&	5.3&			111&29.5 &	19.9&{15.9}\\
 \small{SN\,2009js}&   0.63				&	15					&	11	&		7.3&5.4 &	3.1	&{0.9}\\
 \small{SN\,2011ja}&   	0.21				&	20					&	5.2	&	{1.1}&{0.28} &{1.77}	&{1.6}\\
 \small{SN\,2012cc}&   	0.72				&	15					&10			&	11.5&7.7 &11.6	&{3.7}\\
 \small{SN\,2016adj}& 0.196                 &8.8&1.7&{ 0.24}&{0.05} &{0.25}&{0.20}\\
\hline
  \end{tabular}  
 \caption{Upper limits (ULs) on the integrated flux above the energy threshold and above 1 TeV. {These ULs are computed assuming 95$\%$ {confidence level and a power-law index of 2}}. The associated ULs on the luminosities are computed using the distances reported in {Table}\,\ref{tab:SNcrit}. Upper limits on ${\dot{M }}/{u_\mathrm{w}}$ are derived from methods using the average time and a fit on the LCs, both using equation\,\ref{equation_2} (see text).}
 \label{tab:ul}
\end{table*}

\section{Discussion}
\label{sec:discuss}

The serendipitous nature of the observations provided us with a sample with a {large diversity in distances, post-explosion delay times and observing conditions}. This has to be kept in mind when interpreting the non-detection of TeV emission from these SNe. For instance, two nearby SNe (SN 2008bk, SN 2011ja) {have been observed around 100 days after the discovery} and we might have missed the periods of maximum TeV emission. By contrast, {early and relatively deep observations towards SN 2004cx and SN 2005dn have been performed, but these two SNe are beyond 20 Mpc {in distance}, and this may explain their non-detection}. For SN 2016adj, as already mentioned, the candidate was very promising in terms of distance and {time delay}, although the duration of observations was not as long as initially planned due to bad weather conditions. Despite the diversity in the observed sample, {these H.E.S.S. observations can be used to derive constraints} on a key {parameter impacting} the SN gamma-ray luminosity, {namely} the mass-loss rate of the progenitor star, which determines the CSM density.

The sample consists mostly of type IIP SNe, the most commonly observed type of {cc-}SNe, for which the progenitor is a cool red supergiant (RSG) star, like the progenitor found for SN 2003gd \citep{Smartt:2004}. Several type IIP SNe show evidence for interaction with dense environments in the form of non-thermal radio emission and X-ray emission \citep[e.g.][]{Chevalier:1982b, Pooley:2002}. These progenitors are believed to have mass-loss rates of typically $10^{-6}-10^{-4}$~M$_\odot\ \mathrm{yr}^{-1}$. However, some progenitors 
{appear} {to exhibit} mass-loss rates as high as $10^{-4}-10^{-3}$~M$_\odot\ \mathrm{yr}^{-1}$ \citep{Smith:2014}. {When combined with the relatively slow RSG winds \citep{Chevalier:review}, these mass-loss rates may lead to the right circumstances in terms of density for particle acceleration to proceed up to very high energies and for gamma-ray emission to be potentially detectable} \citep[e.g.][]{Moriya:2011,Marcowith:2014}. Moreover, there is accumulating evidence of enhanced mass-loss rates from progenitors in the last years prior to explosion \citep[e.g.][]{Fuller:2017}, which may similarly lead to enhanced CR proton acceleration and gamma-ray emission.

For type IIb SNe, like SN 2016adj, the mass-loss rate is predicted to be typically over $10^{-5}$~M$_\odot\ \mathrm{yr}^{-1}$ for a wind velocity of ${u_\mathrm{w}}\simeq$ 10 km s$^{-1}$, as {estimated} for SN 1993J \citep{Tatischeff:2009}. However, there is evidence for a sub-type of type IIb SNe depending on the compactness of the progenitor. {A more compact, less luminous progenitor with a lower H mass envelope and a high-speed wind, would produce a lower density environment} \citep[][]{ChevalierSoderberg:2010}.
We will refer to this type of SNe as compact type IIb. Below we will discuss the implications of the obtained H.E.S.S. ULs in terms of the mass-loss rates of the ten observed SNe. 

\subsection{Modelling}
\label{sec:modelling}

To place our flux upper limits into the context of the SN environment, we use a semi-analytical model for {cc-SNe} described in \citet{Dwarkadas:2013}. {The author predicts a gamma-ray flux of hadronic origin} from SNe and young SNRs based on the hydrodynamical evolution described in \citet{Chevalier:1982a} and \citet{ChevalierFransson:1994}, and the gamma-ray emissivity formula prescribed by \citet{Drury:1994}. The model assumes a constant stellar mass-loss rate and wind velocity, usually known as the steady wind scenario \citep{Chevalier:1982a}. Under this assumption, the {CSM density} is given by the continuity of mass equation: $\rho_\mathrm{amb} = {\dot{M }}/{4\pi u_{\mathrm{w}}} r^{-2}$, which shows that a combination of a high mass-loss rate and a low wind velocity will lead to a high-density CSM. Low wind speeds of $\sim$10~$\mathrm{km\ s}^{-1}$ are commonly realised in RSG progenitors \citep[e.g.][]{Smith:2009}. 

For cc-SNe, the model of \citet{Dwarkadas:2013} gives the following relation of the expected gamma-ray flux as a function of stellar mass-loss parameters, SN explosion characteristics and time, $t$, since the explosion:
\begin{equation}
\label{equation_1}
\begin{split}
 F_\mathrm{\gamma}(>\,E_{0} ,\mathrm{t})&=\frac{3q_{\mathrm{\alpha}}\xi(\kappa C_{1})m^{3}}{32\pi^{2}(3m-2)\beta\mu m_\mathrm{p}d^{2}}\left[ \frac{\dot{M }}{ u_\mathrm{w}}\right] ^{2} t^{m-2}.
\end{split}
\end{equation}

{ This equation is valid as long as the maximum photon energy E$_{ph,max}$ (related to the maximum energy of accelerated particles E$_{max}$) is significantly higher than E$_{0}$. As shown by \citet{Marcowith:2018}, in the case of cc SNe evolving in their dense wind progenitor such as SN~1993J, E$_{max}$ conservatively remains above $\sim$0.5 PeV (i.e.~E$_{ph,max}$ $\gtrsim$ 30 TeV) during the first year after the SN, so that equation \ref{equation_1} can be used as is.} {The variable $t$ is the elapsed time in days since the SN explosion, and the variable $d$ is the distance in Mpc as given in Table\,\ref{tab:SNcrit}}. {The variable} $q_{\mathrm{\alpha}}$ is the gamma-ray emissivity normalised to the {hadronic CR energy} density, for which values are tabulated in \cite{Drury:1994}. We use $q_{\mathrm{\alpha}}  (\geq 1 \text{TeV}) = {1.02 \times 10^{-17}}$ s$ ^{-1} $ erg$ ^{-1}$ cm$^{3}$ (H-atom) $^{-1}$, which corresponds to a gamma-ray spectral index of 2, {adopting the value assumed for SN~1993J in \cite{Tatischeff:2009}{. This study clearly shows that the sub shock and total compression ratios both are close to 4, meaning the shock remains weakly modified throughout the SN~1993J time evolution}. }The potential systematic error introduced {by this assumption can be quantified} by considering the extreme case of a steep spectral index of 2.4. In such a case, the gamma-ray emissivity would become $q_{\mathrm{\alpha}} (\geq 1 \text{TeV}) = {8.1 \times 10^{-19}}$ s$ ^{-1} $ erg$ ^{-1}$ cm$^{3}$ (H-atom) $^{-1}$. This would lower the flux values obtained in equation\,\ref{equation_1} by a factor of $\sim$12. 
{The parameter $\xi$ is the fraction of the shock energy flux that {is converted into} CR proton energy, and $\beta$ {is the fraction of the total volume, $\mathcal{V}$, that is already shocked and where the density of target protons is high ($\mathcal{V}_\mathrm{shocked} = \beta4 \pi R_\mathrm{sh}^3/3$).} $m_{\mathrm{p}}$ is the proton mass, and $\mu$ the mean molecular weight of the nuclear targets in the {CSM}. We set $\xi$ equal to 0.1, $\beta$ to 0.5 and $\mu=1.4$ following {the parameters chosen for cc-SNe in} \citet{Dwarkadas:2013}. }
{Finally, the parameter $\kappa$ is the ratio of the forward shock (FS) radius to the contact discontinuity (CD) radius. $C_{1}$ is a constant that can be expressed in terms of the geometry of the explosion, as the radius of the FS is defined as $R_\mathrm{sh}=\kappa R_\mathrm{CD} = \kappa C_{1} t^{m}$. 

For this study, we substitute $\kappa C_{1}$ with $V_\mathrm{sh}/ (m\,t^{m-1})$ where $V_\mathrm{sh}$ is the shock velocity and $m$ is the expansion parameter, leading to the following relation:}
\begin{equation}
\label{equation_1bis}
\begin{split}
F_{\gamma}(\,E_{0} ,t)&=\frac{3q_\mathrm{\alpha}\xi(V_\mathrm{sh})m^{2}}{32\pi^{2}(3m-2)\beta\mu m_\mathrm{p}}\left[ \frac{\dot{M }}{ u_\mathrm{w}}\right]^{2}\left(\frac{1}{d^{2}}\right)\left(\frac{1}{t}\right).
\end{split}
\end{equation}
According to the model of \citet{Chevalier:1982b} for a steady wind scenario, $m$ can be expressed as $m = (n-3)/(n-2)$, where $n$ is the index of the ejecta density profile ($\rho_{\mathrm{ej}} \propto r^{-n}$). For $n$, \citet{Chevalier:1982b} {has found} values between 7 and 12, implying {that $m$ lies} between 0.8 and 0.9, in agreement with observations of some radio SNe \citep[e.g.][]{Weiler:2006}. We adopt here $m=0.85$, and a shock velocity $V_\mathrm{sh}=10,000\,\mathrm{km\,s}^{-1}$ {as fiducial parameters.} 
{The dependence on $1/t$ breaks down for $t\rightarrow 0$, but particle acceleration does not immediately start at $t=0$, as the shock first needs to break out of the star, and some time ({ days-weeks, depending on the B-field value and the turbulence)} should be allowed for the particles to be accelerated to high enough energies to produce VHE gamma rays. Note also that $t=0$ should refer to the time of core collapse, whereas in our analysis we had to use the time of SN detection, which could be several hours or even days after the actual explosion time. Note that within the first week after {core collapse the SN may be so bright in the optical band that the gamma-ray emission is strongly attenuated by gamma-gamma interactions, as} explained in section~\ref{g-gopacity}.
}

Our ULs on the gamma-ray flux above 1 TeV {given in Table \ref{tab:ul}} can be converted into an upper value for ${\dot{M }}/{ u_\mathrm{w}}$ by inverting equation \ref{equation_1bis}, replacing the constant parameters by their {respective values and expressing the mass-loss rate such that $\dot{M}$ = 10$^{-5}$~$\dot{M}_{-5}$~M$_\odot\ \mathrm{yr}^{-1}$} and the wind velocity $u_\mathrm{w}$ = 10 $u_{\mathrm{w},10}$ km\,s$^{-1}$: 
\begin{equation}
\label{equation_2}
\begin{split}
\left[ \frac{\dot{M}_{-5} }{ u_\mathrm{w,10}}\right] ^{2}\leq \frac{F_\mathrm{\gamma}(>1\text{TeV} ) \,d_\mathrm{Mpc}^{2}\,t_\mathrm{day}}{5.14 \times 10^{-12}}.
\end{split}
\end{equation}

The numerical value of 5.14$ \times$ 10$^{-12}$ is of the same order as {that} derived in \citet{Tatischeff:2009} for SN~1993J (within a factor $\lesssim$ 2), and comparable to that obtained by \citet{Murase:2014} {with the same parameters}, within a factor $ \lesssim$ 4.

{To establish the ULs on ${\dot{M}}/{u_\mathrm{w}}$, we use two methods: the first one simply consists in substituting the variable $t$ by the exposure-weighted average time reported in Table \ref{tab:statistics}. A second method consists in fitting with equation\,\ref{equation_2} the nightly binned flux points (cf. Figures \ref{LC_2016adj} and \ref{fig:LightCurves_all}) with their respective dates expressed relatively to the discovery date. For this method, we set the fitting function given in equation~\ref{equation_2} to zero at t~$<$~5 days, in order to account for the possibly strong gamma-ray attenuation through gamma-gamma interactions during the early stages of the SN evolution (see Section~\ref{g-gopacity}). The goodness of the fit is estimated  by a $\chi^2$ test. Fitting the lightcurves in order to constrain the mass-loss parameter is very sensitive to the gamma-ray flux  immediately following the SN explosion, whereas {using the exposure-weighted} time $t$ is more sensitive to the average measured flux. Results and methods are discussed in the next section.}

\subsection{Derived upper limits on the wind properties}

{Upper limits representing a 2$\sigma$ level of the ${\dot{M}}/{u_\mathrm{w}}$ ratio}, derived from the two methods described in the previous section, are given in Table \ref{tab:ul} in units of 10$^{-5}$~$\mathrm{M_{\odot}~yr^{-1}~km^{-1}\,s}$.} 
The constraints on $\dot{M}$, assuming $u_\mathrm{w}$ = 10 km\,s$^{-1}$, are shown in Figure \ref{graph_1}.} 

{In general, more constraining ULs are obtained through the fit to the lightcurves, as compared to the method using the time-average flux limits: this is the case for SN 2004cx, and SN 2005dn, for which fluxes are determined shortly after the SN explosion dates, as well as as observation spanning the whole year. 
For 2005dn, the value obtained with the fit method seem to favor the first sets of flux points mostly negatives with small errors, compared to late observations taken after a gap which are showing more positive values. The same case seems to happen for SN 2008bk, for which observations were all taken $\sim$ 100 days after the discovery but are spanning over several months. For SNe observed only during a short time span, like SN 2009hf, SN 2011ja and SN 2016adj, the two methods give similar ULs, as expected. The method using a fit to equation \ref{equation_2} gives more weight to the early observations, as this is where the highest fluxes are expected and relies on the assumption that the gamma-ray flux evolution follows exactly the $1/t$ scaling. In reality, the progenitor mass loss history may be more complicated, and there is also some uncertainty regarding the onset of particle acceleration and the effect of gamma-gamma attenuation. As such, the time-average flux method gives perhaps a less precise but more conservative constraints on $\dot{M}/u_\mathrm{w}$.} 

For the nine SNe of the serendipitous sample, {both methods give ULs lying between $\sim$ 2.0 $\times$ 10$^{-5}$~M$_\odot\ \mathrm{yr}^{-1}$ and $\sim$ 2.0 $\times$ 10$^{-3}$~M$_\odot\ \mathrm{yr}^{-1}$ }, and are consistent with predictions for type IIP SNe with a RSG progenitor having a mass-loss rate in the range $10^{-6} - 10^{-4}$~M$_\odot\ \mathrm{yr}^{-1}$. These constraints { {show} large values in $\dot{M}/u_\mathrm{w}$, but still within the range of expected mass-loss rates for some RSGs. }

{For SN 2016adj, the mass-loss rate UL, {confirmed by both methods}, is reaching $\sim$ 2.5 $\times10^{-5}$ M$_\odot\ \mathrm{yr}^{-1}$, as the SN {occurred in the nearby Cen A galaxy and} has been observed very early: it is of the same order as the value estimated for the well-studied SN 1993J \citep[$\mathrm{\dot{M}}$ = 3.5 $\times$ 10$^{-5}$~M$_\odot\ \mathrm{yr}^{-1}$ for $u_w = 10 $ km\,s$^{-1}$,][]{Tatischeff:2009}, both SNe being of Type IIb occurring at similar distances}. We note that the expansion parameter of SN\,1993J near the time of discovery is estimated to be high, $m = 0.919 \pm 0.09$, as implied by early radio observations \citep{Bartel:2002}. {Apart from this difference, given that the TeV gamma-ray flux from SN 1993J was predicted to be at the level of sensitivity of current IACTs \citep{Marcowith:2014}, TeV emission from SN 2016adj could have been detected with H.E.S.S. if it were to share the same environmental properties as SN 1993J.}

\begin{center}
\begin{figure}
\includegraphics[width=0.99\textwidth]{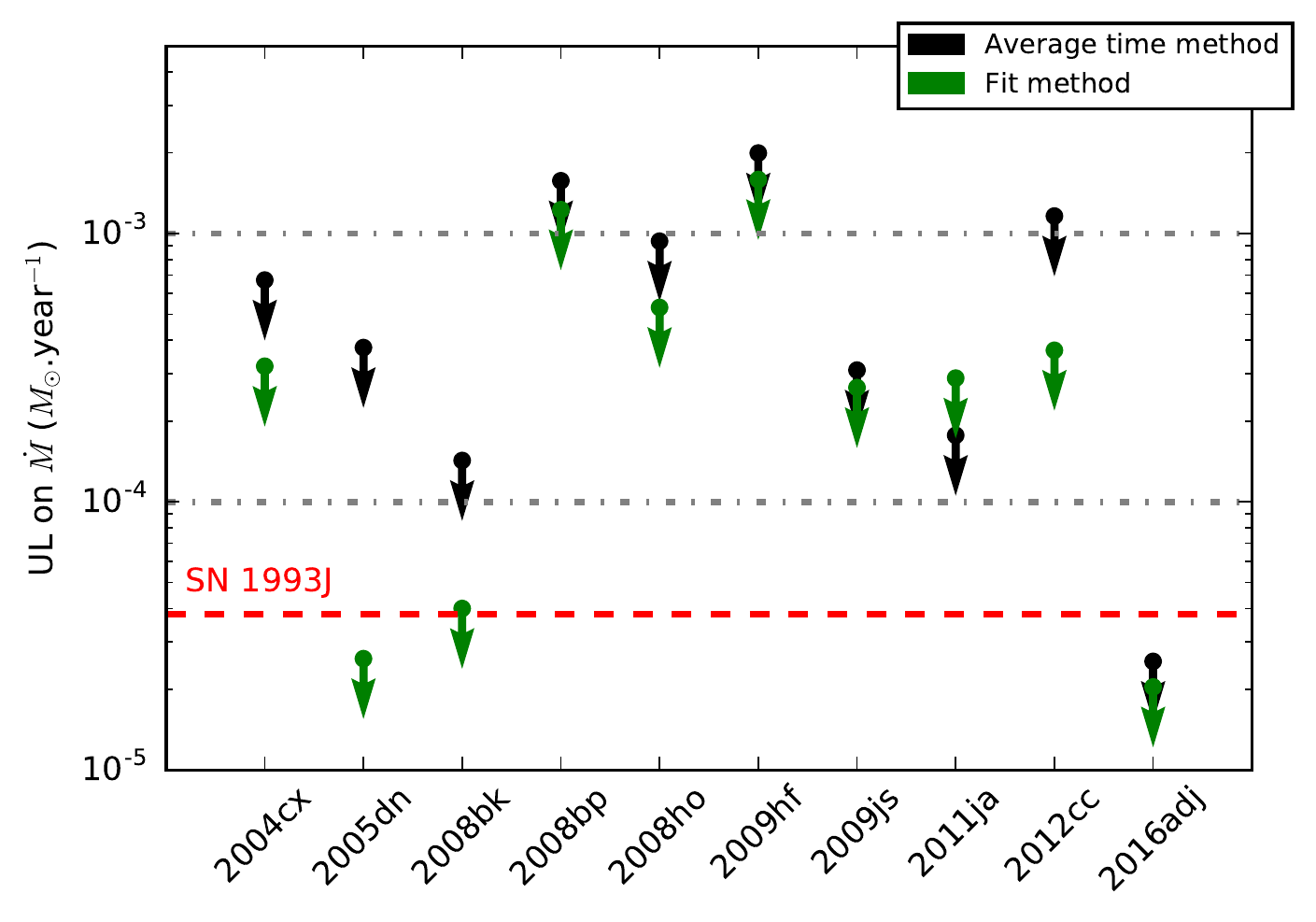} 
\caption{Upper limits on {progenitor mass-loss rates, $\dot{M}$,} assuming $u_\mathrm{w}$ = 10 km\,s$^{-1}$ for the ten cc-SNe investigated in this study, {derived with two methods (see text).} The mass-loss {rate for SN\,1993J, derived by \citet{Tatischeff:2009}, is} also shown.}
\label{graph_1}
\end{figure}
\end{center}

\subsection{{Opacity} due to gamma-gamma absorption.}
\label{g-gopacity}
At short timescales after the explosion, when the SN {is near maximum optical peak luminosity}, the VHE gamma-ray emission may be suppressed due to electron-positron pair-production, arising from the interaction of TeV photons with low-energy photons from the SN photosphere.{ This gamma-gamma absorption critically depends on the photosphere properties of each object, and no complete model exists so far. First attempt to quantify this effect in SN~1993J has been carried out by \citet{Tatischeff:2009} under the assumption of isotropic interactions with photospheric photons. More recently, preliminary calculations of this time-dependent absorption accounting for the geometrical effects have been performed by \citet{Marcowith:2014} in the case of SN~1993J for which the parameters of the SN hydrodynamical evolution and photosphere are well known. While the absorption is expected to be very large during the first week, the attenuation amounts to a factor of a few at t $\sim$ 10-20 days before gradually decreasing at the level of $\sim$ 10-20\% on a month timescale.}
{ Thus}, pair-production may have an impact on the detectability of gamma-ray emission from the SNe in the first week after the explosion, depending on the luminosity evolution of a given SN. Given the large spread in the distribution of time delays between the H.E.S.S. observations and the SN discoveries for the SN serendipitous sample, this opacity can, in general, be ignored. 
        
The only object for which this attenuation should be considered is SN~2016adj, whose ToO observations are clustered at short time delays. {If the attenuation of TeV photons 
is at a similar level as {that estimated in} SN~1993J}, the expected VHE gamma-ray flux would be much lower during a large part of the H.E.S.S. observation time window and this would explain the non-detection of any excess towards SN~2016adj. In other words, constraints on the VHE gamma-ray flux could not be directly translated into limits on the mass-loss rate until the above-mentioned parameters regarding the evolution of the SN shock and photosphere are better known. {SN 2016adj occurred very close to the center of Cen A, so it may well be that the local environment led to an additional gamma-gamma absorption and this could also explain the non-detection.}

\subsection{Discussion on the SNe environment in perspective of other observations.}
{{There may be another explanation for} the non-detection of SN 2016adj, for which the ULs are near or below the predicted gamma-ray flux, namely that this interesting candidate does not present the right {environmental} properties. This could be the case if SN 2016adj was a compact type IIb SN (cIIb), as opposed to SN 1993J, which is known to have been an extended type IIb \citep[eIIb,][]{ChevalierSoderberg:2010}}. For type cIIb SNe, the wind velocity is probably higher (${ u_\mathrm{w}} \gtrsim$ 100 km s$^{-1}$), {implying a lower density CSM. Under the hypothesis that SN~2016adj is of type cIIb, assuming $u_\mathrm{w}$ = 100 km\,s$^{-1}$, the upper limit on the mass-loss rate would increase by a factor of 10 (see equation \ref{equation_2}), i.e.~$\mathrm{\dot{M}} <$\,3.0 $\times$ 10$^{-4}$~M$_\odot\ \mathrm{yr}^{-1}$, and could better accommodate the H.E.S.S. non-detection of this {apparent} SN~1993J~twin.} {\citet{ATel1302} proposed that SN~2016adj originated from a progenitor with a lower luminosity than the SN 1993J progenitor, suggesting that the explosion might be of compact IIb-type. This claim, however, was later disfavoured by \citet{ATel8759}.}
{Further multi-wavelength observations will weigh in on the nature of this object, and an} analysis of radio and X-ray data is forthcoming (Hajela at al. in prep.). 
{For the rest of the sample, {in addition of} being Type II SNe (mostly IIP), the nine selected candidates} did not have any strong {indication} of the conditions necessary for significant early CR acceleration. Several of the SNe have been studied at various levels of detail, with results consistent with our study.
{Radio and X-ray observations are of prime interest to check the consistency of our result, as they can provide additional constraints on the wind parameters.
{The literature reports radio and X-ray observations for only one SNe in our sample}, namely for the rather close-by object SN\,2011ja.} \citet{Chakraborti:2013} suggest that {the measured SN\,2011ja radio and X-ray fluxes} are consistent with an expansion into a low-density bubble and interaction with an inhomogeneous circumstellar medium formed by a RSG ($\sim$12\,M$_{\odot}$). The corresponding progenitor mass-loss rate  was estimated to be of the order of $\sim$10$^{-6}$\,M$_\odot\ \mathrm{yr}^{-1}$, consistent with the {constraint} of $<$\,1.6 $\times$ 10$^{-4}$\,M$_\odot\ \mathrm{yr}^{-1}$ derived {from our non-detection at TeV gamma rays.}

SN\,2008bk was observed to have a progenitor mass of $\sim$8\,M$_{\odot}$ and {a luminosity of $\log{(L/L_{\odot})}\sim 4.5$} \citep{Davies:2017} after its post-explosion disappearance was confirmed by \citet{Mattila:2013} and \citet{vanDyk:2013}. By applying the \citet{deJager:1988} prescription \citep[see also][]{Mauron:2011}, the observed luminosity implies a pre-SN progenitor mass-loss rate of $\sim$10$^{-6.3}$\,M$_\odot\ \mathrm{yr}^{-1}$, consistent with our model-dependent constraint of $<$1.4 $\times 10^{-4}$\,M$_\odot\ \mathrm{yr}^{-1}$.

We further {mention a study of the spectral evolution of 122 nearby SNe including} post-explosion spectral line observations of SN\,2008bp at wavelengths 4000-9500\,\r{A} at 12, 40 and 47 days \citep{Gutierrez:2017}. The authors noted SN\,2008bp to be the only event to not have Fe-group and H$_\mathrm{\gamma}$ line-blending in early stages of evolution. This characteristic might suggest that the circumstellar medium of SN\,2008bp was less dense than that of the rest of the sample. 
Outlier behaviour was also discovered for SN\,2009js. \citet{Gandhi:2013} found the event to be subluminous, suggesting a low ejecta mass and explosion energy. 
It follows that the environments of SN\,2008bp and SN\,2009js are not likely conducive to the TeV gamma-ray detection sought in our study, but are consistent with the mass-loss rate upper-limits of $\sim$1.6 $\times$ 10$^{-3}$ and $\sim$3.1 $\times$ 10$^{-4}$\,M$_\odot\ \mathrm{yr}^{-1}$, respectively, as derived from the H.E.S.S. observations. 
\begin{center}
\begin{figure*}
\includegraphics[width=0.9\textwidth]{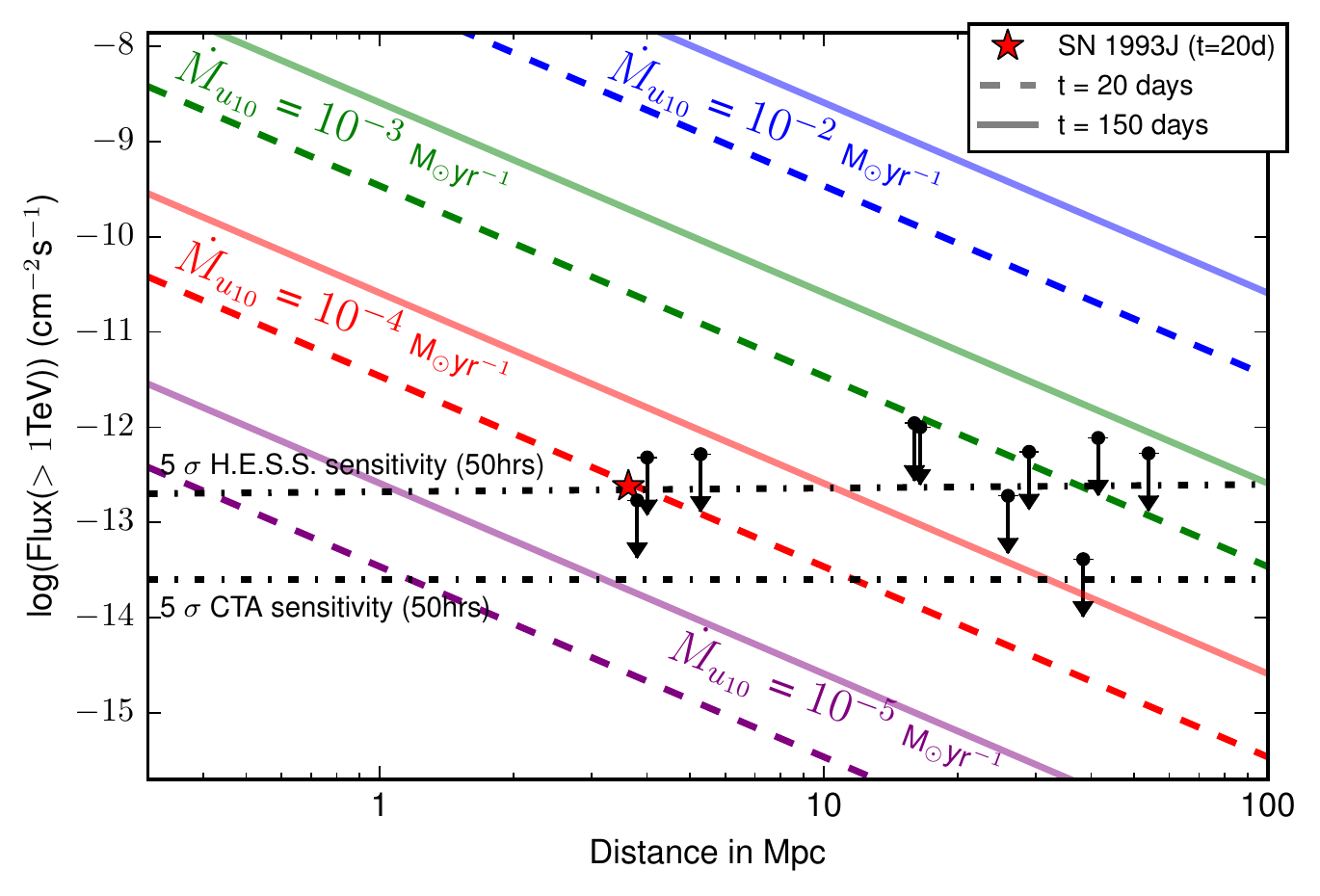} 
\caption{Predicted flux above 1 TeV using equation \ref{equation_1} as a function of the distance to the source. Mass-loss rates are {given in units of} M$_\odot\ \mathrm{yr}^{-1}$ {assuming} $u_\mathrm{w}=10$\,km s$^{-1}$ and the parameters described in {Section}\,\ref{sec:modelling}, for $t$ = 20 days ({solid} lines) and $t$ = 150 days (dashed lines) after the SN explosion. {The Mass-loss rate is given} for each pair of lines of the same color. The expected flux for SN 1993J is computed using equation \ref{equation_1} and $t=20$ days. {The CTA sensitivity for a 50 hr long observation is taken from \citet{CTA:2013}. {The ten ULs on the flux above 1 TeV derived in this study are also shown (see text).}}}
\label{graph_2}
\end{figure*}
\end{center}

\subsection{Prospects for future observations}
In order to put {the flux ULs derived in this work} in perspective, Figure \ref{graph_2} shows {the gamma-ray flux computed with equation\,\ref{equation_1bis} as a function of the distance for different values of the pre-SN mass-loss rate, together with the typical { five sigma }point-source sensitivities of H.E.S.S. and of the next generation of IACTs, {the} Cherenkov Telescope Array \citep[CTA,][]{CTA:2013}. For comparison our upper limits are shown, which, as expected, {are close to the 50~h H.E.S.S. sensitivity, $\simeq 2 \times 10^{-13}$\, cm$^{-2}$s$^{-1}$ \citep{Aharonian:2006}, bearing in mind the exposure times.} Parameters are chosen as described in Section\,\ref{sec:modelling}, with $q_{\mathrm{\alpha}}$  {corresponding} to a {flat spectral index}. Two different values for the {time delay since the SN explosion} are considered: $t=150$~{days} corresponds to a value that is {representative of} our sample, whereas $t=20$~{days} {roughly corresponds to} the optical peak luminosity of a SN. {As seen in Figure \ref{graph_2}, the VHE gamma-ray flux at $t=20$ {days} from a SN~1993J-like event is within reach with current IACTs like H.E.S.S., and would clearly be detected by CTA. At $t=150$ {days}, and for mass-loss rates higher than 10$^{-4}$~M$_\odot\ \mathrm{yr}^{-1}$, CTA may detect cc-SNe as far out as the Virgo cluster (16~Mpc).}

Mass-loss rates above $10^{-4}$~M$_\odot\ \mathrm{yr}^{-1}$ are not uncommon, but are usually confined to Type IIb and IIn SNe. Each of these types represent, respectively, $\sim$10\% and $\sim$8 \% of the total cc-SN rates according to \citet{Smith:2011}. The number of cc-SNe occurring in a year can roughly be expressed as a function of the total available stellar mass in 10$^{10}$ M$_{\odot}$ units. {In our Galaxy, this implies a cc-SN rate of about {two} per century \citep[see e.g.][]{Lietal:2011}}. A galaxy cluster in the local Universe, as the Virgo cluster, has a total stellar mass of the order of 10$^{13}$ M$_{\odot}$ \citep[see e.g.][]{O'Sullivan:2017}, which would bring the number of cc-SNe up to $\sim$10 per year in a near radius of 16\,Mpc. This number is very similar to the number of objects predicted by \cite{Horiuchi:2011} for a volume of radius $\lesssim$10\,Mpc. Another example is given by \cite{Smartt:2009}, who identified 5 type IIP SNe in 1999 within $\sim$18\,Mpc, which may represent 60-70$\%$ \citep[see e.g.][]{Smartt:2009, Smith:2011} of the whole sample of cc-SNe of that year. Note that type II objects can also exhibit enhanced pre-SNe mass-loss rates above 10$^{-3}$~M$_\odot\ \mathrm{yr}^{-1}$ \citep[e.g.][]{AlakRay:2017,Arcavi:2017}, and other studies {have shown} that such high mass-loss rates are not so rare among type IIb SNe \citep{Fuller:2017,Ouchi:2017}.
It is then reasonable to expect $\sim$1 to 2 cc-SNe with $\dot{M}$ > $10^{-4}$~M$_\odot\ \mathrm{yr}^{-1}$ occurring per year, within a radius of 18\,Mpc, whatever the sub-type. Such nearby cc-SN events offer a great opportunity for the detection of gamma rays using IACT observations triggered by observations at optical wavelengths.{ Such a ToO program to observe cc-SNe as distant as 10 Mpc is currently in place within the H.E.S.S. collaboration.} { For the expected gamma-ray luminosities of supernovae, the wide-field TeV observatories as HAWC \citep{HAWC:2013} and in the future LHAASO \cite{Vernetto:2016} are less ideally positioned for detecting gamma-ray emission below 100 TeV, as they require relatively long integration times of up to a year to reach the required sensitivity, whereas the gamma-ray flux is declining on shorter time scales. However, their all-sky monitoring capabilities could lead to early detection of unexpectedly bright gamma-ray SNe events.}

\section{Conclusion}

We selected a sample of nine type II SNe that were observed by chance with H.E.S.S. within one year after the SN event, and in addition we triggered ToO observations on SN 2016adj. No significant gamma-ray signal has been detected from any of {these ten} SNe and we derived flux upper limits of the order of $10^{-13}\, $TeV cm$^{-2}$ s$^{-1}$. 

This result is amending previous efforts \citep{Lennarz:2013} and {complements other} recent non-detections, namely the Fermi-LAT studies of type IIn SNe at GeV energies \citep{Ackermann:2015}, the upper limit at TeV energies established by the MAGIC collaboration for the closest type Ia SN 2014J \citep{Ahnen:2017}, and a recent search for GeV emission from super luminous SNe using Fermi-LAT data by \citet{Renault-Tinacci:2017}. {Concerning SN 2016adj, the H.E.S.S. UL is the first constraint derived on this nearby SN event in the gamma-ray domain.}
{The lack of gamma-ray detection reported here, however, does not necessarily indicate that the early phase of SN evolution is not generally conducive to CR acceleration. Instead, the non-detection suggests that it does not occur in this subset of the SNe, which have CSM that are not likely to be dense enough for particle acceleration.}

Using the model developed in \cite{Dwarkadas:2013}, {we expressed our ULs} in terms of constraints on the mass-loss rates of the {SN progenitors}, which turn out to be a few times higher {than, or of the same order as,} the estimated mass-loss rate for the close-by radio-bright SN 1993J.
With the same model, we {predicted} that objects with a mass-loss rate of the order of $10^{-4}$~M$_\odot\ \mathrm{yr}^{-1}$ and distance of $\sim$10\,Mpc could be {detected} very early after the outburst by the current generation of telescopes and a fortiori by the next generation, namely the Cherenkov Telescope Array, CTA \citep{CTA:2013}. 
{In our study, we did not observe candidates with the {required} properties for the detection of gamma rays with H.E.S.S., but our model-dependent investigation suggests that core-collapse SNe will be detected by Cherenkov arrays in the future.}

\section*{Acknowledgements} 

{The authors would like to thank R. Margutti for her useful input as well as the anonymous referee for his/her constructive comments}. R. Simoni thanks M. Renzo and M. Zapartas for helpful discussions. The support of the Namibian authorities and of the University of Namibia in facilitating the construction and operation of H.E.S.S. is gratefully acknowledged, as is the support by the German Ministry for Education and Research (BMBF), the Max Planck Society, the German Research Foundation (DFG), the Helmholtz Association, the Alexander von Humboldt Foundation, the French Ministry of Higher Education, Research and Innovation, the Centre National de la Recherche Scientifique (CNRS/IN2P3 and CNRS/INSU), the Commissariat à l’énergie atomique et aux énergies alternatives (CEA), the U.K. Science and Technology Facilities Council (STFC), the Knut and Alice Wallenberg Foundation, the National Science Centre, Poland grant no.2016/22/M/ST9/00382, the South African Department of Science and Technology and National Research Foundation, the University of Namibia, the National Commission on Research, Science $\&$ Technology of Namibia (NCRST), the Austrian Federal Ministry of Education, Science and Research and the Austrian Science Fund (FWF), the Australian Research Council (ARC), the Japan Society for the Promotion of Science and by the University of Amsterdam. We appreciate the excellent work of the technical support staff in Berlin, Zeuthen, Heidelberg, Palaiseau, Paris, Saclay, Tübingen, and in Namibia in the construction and operation of the equipment. This work benefitted from services provided by the H.E.S.S. Virtual Organisation, supported by the national resource providers of the EGI Federation. 
\bibliographystyle{aa}%
\bibliography{ReferencesSNe}
\appendix
\section{Statistical analysis}
\begin{table}[ht]
\begin{center}
\begin{tabular}{|l|l|c|c|c|c|c|c|c|c|}
\hline
SNe		& N$_\mathrm{on}$ & N$_\mathrm{off}$ &$ \alpha$ & N$_\mathrm{excess}$		&	Sig 	& { Livetime} & Obs. coverage 	& {Average time} \TBstrut\\ 		
			& &&&&&(hrs) &(days) &(days)\\
\hline
SN\,2004cx	&169 &10387&	0.015&	  8.7			&		0.7			&40&-6 - 359&	180\Tstrut\\
SN\,2005dn	&571 &11452&	0.053&		-39		&		-1.5			&53&-3 - 364&	120\\
SN\,2008bk	&50 &3652&	0.018&		-18		&		-2.3			&9.6&98 -	211& 136\\
SN\,2008bp	&32 &1860&	0.017&		1.1			&		0.2			&4.7&272 - 282&	282\\
SN\,2008ho	&9 &369	&	0.030&	 	-2.3			&		-0.7			&1.4&34 - 	34&34\\
SN\,2009hf	&43 &1404&0.029&		3.3			&		0.5			&4.0&124 - 134 &133\\
SN\,2009js 	&14 &711&	0.015&			3.4		&			1		&4.8&1 - 35&17.5\\
SN\,2011ja	    &37 &620& 0.053&		4.51		&		0.75			&3.4&91 - 131&111\\
SN\,2012cc 	&7&660&	0.013&				-1.9		&		-0.7			&3.0&53 - 343 &255	\\
SN 2016adj &624&8573&0.070&22&0.9&13 &3 - 10& {7} \\
\hline
\end{tabular}
\caption{Observed statistics for each SN (see text). Sig stands for significance and the observation range gives the number of days {since the SN} discovery date for the first - last observation run. The average time is the exposure-weighted mean time in days since the discovery date. }  \label{tab:statistics}
\end{center}
\end{table}
\section{Lightcurves}

\label{app:lightcurves}
\begin{figure*}[ht]
\begin{center}
\includegraphics[width=0.9\textwidth]{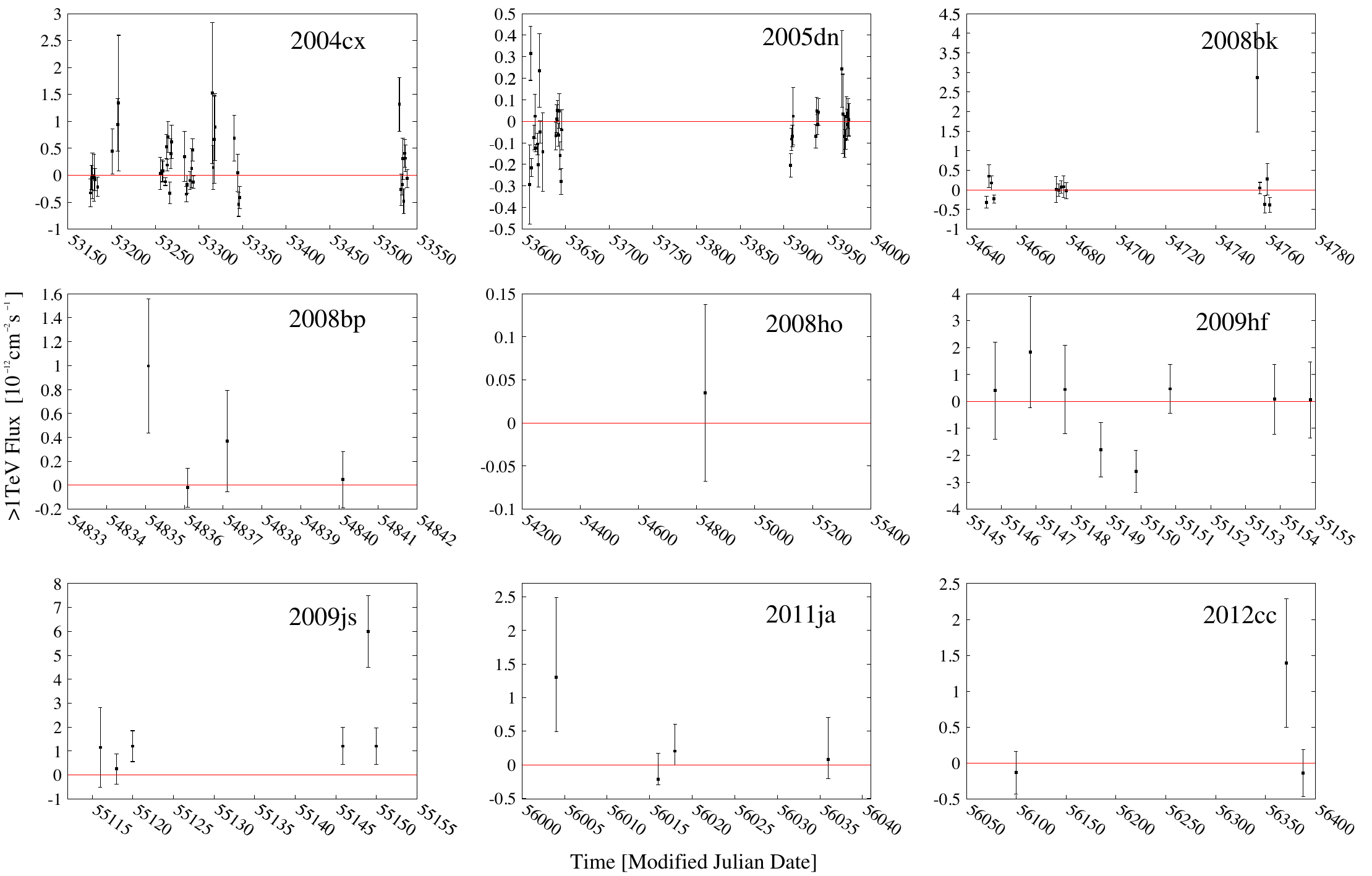}
\caption{Time evolution binned on a nightly basis of the integrated flux above 1 TeV as measured with H.E.S.S. towards the nine {serendipitously observed} SNe considered in this study. {The red line outlines the zero level.}}
\label{fig:LightCurves_all}
\end{center}
\end{figure*}
\end{document}